\documentclass[letterpaper, 10 pt, conference]{ieeeconf}  

\IEEEoverridecommandlockouts                              

\usepackage{amsmath,amssymb,amsfonts,mathrsfs,dsfont}
\usepackage{algorithm, algpseudocode} 
\usepackage{graphicx}
\let\labelindent\relax
\usepackage{enumitem} 
\setlist[itemize,enumerate]{leftmargin=*, noitemsep, topsep=0pt}
\usepackage[hidelinks]{hyperref} 

\newtheorem{theorem}{Theorem}[section]
\newtheorem{lemma}[theorem]{Lemma}
\newtheorem{proposition}[theorem]{Proposition}
\newtheorem{corollary}[theorem]{Corollary}
\newtheorem{remark}[theorem]{Remark}

\makeatletter
\def\@opargbegintheorem#1#2#3{%
  \@IEEEtmpitemindent\itemindent\topsep 0pt\rmfamily\trivlist%
  \item[\hskip\labelsep{\indent\itshape #1\ #2}]%
  \itemindent\@IEEEtmpitemindent
  \textit{(#3):}\ \ignorespaces}
\makeatother

\title{\LARGE \bf
Physics-Informed Learning for Multiplayer Differential Games
}

\author{Takuma Adams$^{1}$, Andrew C.~Cullen$^{2}$, and Tansu Alpcan$^{3}$
\thanks{This research was supported in part by the Commonwealth through an Australian Government Research Training Program Scholarship.}
\thanks{$^{1}$Takuma Adams is with the Department of Electrical and Electronic Engineering, The University of Melbourne, Parkville, Australia; and Defence Science and Technology Group, Australia
        {\tt takuma.adams@student.unimelb.edu.au}}%
\thanks{$^{2}$Andrew C.~Cullen is with the School of Computing and Information Systems, The University of Melbourne, Parkville, Australia
        {\tt andrew.cullen@unimelb.edu.au}}%
\thanks{$^{3}$Tansu Alpcan is with the Department of Electrical and Electronic Engineering, The University of Melbourne, Parkville, Australia
        {\tt tansu.alpcan@unimelb.edu.au}}%
}
\usepackage{eso-pic}
\AddToShipoutPictureFG*{\AtPageUpperLeft{\put(54,-18){\parbox[t]{504pt}{\fontsize{7}{8}\selectfont 
\copyright~2026 IEEE. Personal use of this material is permitted. Permission from IEEE must be obtained for all other uses, in any current or future media, including reprinting/republishing this material for advertising or promotional purposes, creating new collective works, for resale or redistribution to servers or lists, or reuse of any copyrighted component of this work in other works. Accepted version of the article published in \emph{IEEE Control Systems Letters}, vol.~10, pp.~619--624, 2026, DOI: \href{https://doi.org/10.1109/Interactive Display}{10.1109/LCSYS.2026.3700693}. The contents of this paper were also selected by the CDC 2026 Program Committee for presentation at the Conference.}}}}

\begin{document}

\maketitle
\thispagestyle{empty}
\pagestyle{empty}

\begin{abstract}
Computing per-player feedback strategies in $N$-player adversarial differential games is challenging as the associated coupled Hamilton-Jacobi-Bellman system is high-dimensional and generally lacks classical solutions. We address this by approximating viscosity solutions using physics-informed neural networks within a policy iteration scheme with a vanishing-viscosity annealing schedule. Theoretically, we establish an explicit $\varepsilon$-Nash guarantee by bounding per-player regret via computable residuals and prove our method reaches this stopping criterion in finite time. For two-player zero-sum games, we further establish a computable bound on the approximation error of the value function. Numerical experiments on pursuit-evasion scenarios illustrate the tightness of these computed bounds and the framework's scalability to high-dimensional environments.
\end{abstract}

\section{Introduction}\label{sec:introduction}
Computing optimal feedback strategies for multi-agent adversarial systems requires solving for a Nash equilibrium governed by coupled Hamilton-Jacobi-Bellman (HJB) equations~\cite{basar_dynamic_1999}. While these partial differential equations (PDEs) encode the necessary conditions for optimality, numerical solutions notoriously suffer from the curse of dimensionality. Classical grid-based solvers become computationally intractable for high-dimensional joint state spaces~\cite{e_algorithms_2022}, with conventional game- and control-theoretic approaches resorting to coarse discretizations or simplified dynamics that fail to capture complex coupled interactions~\cite{cheng_multiplayer_2023, adams_suboptimality_2025}.

Physics-informed neural networks (PINNs)~\cite{raissi_physics_2019} offer an alternative path to resolving this curse of dimensionality, having been successfully applied to single-agent optimal control~\cite{lee_hamiltonjacobi_2025,tayal_physics-informed_2025}, two-player zero-sum~\cite{wadi_trajectory-informed_2025}, and two-player nonzero-sum games~\cite{zhang_parametric_2025, zhang_pontryagin_2024, ghimire_solving_2024, adams_poster_2026}. By approximating value functions via neural networks trained to directly minimize the PDE residuals, PINNs offer a mesh-free paradigm. This aligns learning with the underlying game dynamics, preserving the structural consistency and provable guarantees of the Hamilton-Jacobi formulation---properties unmatched by purely data-driven mesh-free approaches like reinforcement learning (RL)~\cite{orr_multi-agent_2023}.

Despite this promise, a critical gap remains in the theory of PINNs for multiplayer games, as existing results are empirical and often rely on auxiliary supervision to handle discontinuities~\cite{zhang_value_2024}. Rigorous theoretical analysis remains limited to single-agent control~\cite{shilova_learning_2024, kim_hamilton-jacobi-bellman_2020, meng_physics-informed_2024}, with a recent extension to the special case of two-player zero-sum games~\cite{wadi_trajectory-informed_2025}. Notably, while~\cite{kim_hamilton-jacobi-bellman_2020} analyzed PINNs for control using value iteration (VI) and~\cite{wadi_trajectory-informed_2025, meng_physics-informed_2024} provided analysis for policy iteration (PI), extensions of these results to coupled $N$-player adversarial differential games have yet to be established.

We bridge this gap with a PINN-based PI framework for computing per-player optimal feedback strategies in $N$-player adversarial differential games. While recent literature explores PI for two-player zero-sum games~\cite{wadi_trajectory-informed_2025}, we extend the analysis to $N$-player equilibria, preserving stability guarantees from~\cite{meng_physics-informed_2024}. Crucially, we achieve this by adapting existing neural network verification tools, such as $\partial$-CROWN~\cite{eiras_efficient_2024}, to \emph{certify} our PINN residuals. To date, such methods---developed by the adversarial machine learning community to guarantee performance under adversarial manipulation~\cite{cullen_et_2024, cullen2022double}---have not been applied to analyze differential game equilibria. We refer readers to~\cite{li_sok_2023} for a comprehensive review. Our main contributions are
\begin{enumerate}
    \item \textbf{Computable $\varepsilon$-Nash guarantees}: We establish a mathematical framework linking network bounds to per-player regret and $\varepsilon$-Nash equilibria, proving that our algorithm satisfies the resulting stopping criterion in finite time.
    \item \textbf{Value approximation bounds}: We establish a theoretical bound on the pointwise approximation error of the learned value function for two-player zero-sum games.
    \item \textbf{$N$-player physics-informed PI}: We extend PI to $N$-player games via viscous PDEs and annealing, ensuring stability through analytical best-response updates. 
\end{enumerate}
These results are empirically validated on two pursuit-evasion scenarios, demonstrating both tightness in low dimensions and scalability to high-dimensional adversarial environments.

\section{Problem Formulation \& Preliminaries}\label{sec:problem_formulation}
Consider a differential game in which $N$ players are partitioned into two opposing teams, $\mathcal{N}_p$ and $\mathcal{N}_e$, such that $\mathcal{N} = \mathcal{N}_p \cup \mathcal{N}_e$. The joint state $\mathbf{x} \in \mathcal{X} \subset \mathbb{R}^n$ evolves over a finite time horizon $t \in [t_0, T]$ according to the known control-affine dynamics
\begin{equation}\label{eq:generic_state_equation}
     \dot{\mathbf{x}} = f(t, \mathbf{x}(t)) + \sum_{i \in \mathcal{N}} g_i(t, \mathbf{x}(t))\mathbf{u}_i(t)  \,,
\end{equation}
subject to $\mathbf{x}(t_0) = \mathbf{x}_0$, where the drift $f:\Omega \rightarrow \mathbb{R}^{n}$ and input matrices $g_i: \Omega \rightarrow \mathbb{R}^{n \times m_i}$ are assumed Lipschitz continuous on the compact spatio-temporal domain $\Omega = [t_0, T] \times \mathcal{X}$, guaranteeing a unique solution for any set of admissible controls. The actions $\mathbf{u}_i(t)$ are restricted to compact and convex sets $\mathcal{U}_i \subset \mathbb{R}^{m_i}$. We consider a feedback information pattern such that players select actions via $\pi_i: \Omega \rightarrow \mathcal{U}_i$ where $\mathbf{u}_i(t) = \pi_i(t, \mathbf{x}(t))$ and $\Pi_i$ denotes the set of admissible strategies for player $i$. For a joint profile $\pi = (\pi_1 \; \dots \;\pi_N)$, the cost-to-go functional for player $i$ is defined as 
\begin{multline}\label{eq:generic_cost_to_go} 
J_i(t, \mathbf{x}, \pi_i, \pi_{-i}) = \int_t^T r_i (\tau, \mathbf{x}(\tau), \mathbf{u}_i(\tau), \mathbf{u}_{-i}(\tau))d\tau \\
+ \psi_i(\mathbf{x}(T)) \,, 
\end{multline} 
subject to (\ref{eq:generic_state_equation}). Here, the terminal costs $\psi_i$ are Lipschitz continuous, and the running cost $r_i$ is strictly convex in $\mathbf{u}_i$, taking the quadratic form $r_i = \mathbf{u}_i^\top \mathbf{R}_i(t, \mathbf{x}) \mathbf{u}_i + q_i(t, \mathbf{x}, \mathbf{u}_{-i})$ with positive definite $\mathbf{R}_i$.

\subsection{Coupled HJB System \& Viscosity Solutions}
We employ the state-feedback Nash equilibrium solution concept involving strategy profiles $\pi^* = (\pi_1^* \; \dots \; \pi_N^*)$ where $J_i(t, \mathbf{x}, \pi_i^*, \pi_{-i}^*) \leq J_i(t, \mathbf{x}, \pi_i', \pi_{-i}^*)$ holds for all admissible strategies $\pi_i' \in \Pi_i$, $(t, \mathbf{x}) \in \Omega$, and $i \in \mathcal{N}$~\cite{basar_dynamic_1999}. The associated value functions $V_i: \Omega \rightarrow \mathbb{R}$ represent the optimal cost-to-go for player $i$ from any state $(t, \mathbf{x}) \in \Omega$, and can be numerically approximated up to a residual and optimization error, making $\varepsilon$-Nash equilibria the appropriate relaxed solution concept.

A fundamental challenge in $N$-player differential games is that $V_i$ is generally not continuously differentiable ($C^1(\Omega)$). Consequently, the classical HJB equations may not hold in the strict sense. We therefore seek viscosity solutions $V_i$ to the $N$ coupled HJB equations
\begin{equation}\label{eq:HJB}
    \partial_t V_i+ \inf_{\mathbf{u}_i \in \mathcal{U}_i}\Big[ H_i(t, \mathbf{x}, \nabla_\mathbf{x} V_i, \mathbf{u}_i, \mathbf{u}_{-i}^*) \Big] = 0 \,, 
\end{equation}
subject to the terminal boundary condition $V_i(T, \mathbf{x}) = \psi_i(\mathbf{x})$. Given the control-affine dynamics and quadratic cost structure, the Hamiltonian $H_i$ is defined as
\begin{equation}\label{eq:hamiltonian}
    H_i(t, \mathbf{x}, \mathbf{p}_i, \mathbf{u}_i, \mathbf{u}_{-i}) = \mathbf{p}_i^\top \big( f(\cdot) + \sum_{j \in \mathcal{N}} g_j(\cdot)\mathbf{u}_j \big) + r_i(\cdot) \,,
\end{equation}
where $\mathbf{p}_i = \nabla_{\mathbf{x}} V_i \in \mathbb{R}^n$ denotes the costate. To facilitate numerical approximations, we define the Hamiltonian for a fixed costate $\mathbf{p}_i$ and opponent profile $\hat{\mathbf{u}}_{-i}$ as $\Phi_i(t, \mathbf{x}, \mathbf{u}_i; \mathbf{p}_i) = H_i(t, \mathbf{x}, \mathbf{p}_i, \mathbf{u}_i, \hat{\mathbf{u}}_{-i})$. The strict convexity of $r_i(\cdot)$ ensures the best-response action is uniquely given by
\begin{equation}\label{eq:best_response}
     \mathbf{u}_i^{br}(t, \mathbf{x};\mathbf{p}_i) = \arg\min_{\mathbf{u}_i \in \mathcal{U}_i} \Phi_i(t, \mathbf{x}, \mathbf{u}_i; \mathbf{p}_i) \,.
\end{equation}
For notational brevity, we may write $\Phi_i(\mathbf{u}_i; \mathbf{p})$ and $\mathbf{u}_i^{br}(\mathbf{p})$ when $i \in \mathcal{N}$ and $(t, \mathbf{x}) \in \Omega$ are fixed. 

\section{Solving PINN-Coupled HJB Equations}\label{sec:PINNs}
We approximate the $N$ value functions $V_i$ using PINN-trained feedforward neural networks $\hat{V}_i(t, \mathbf{x}; \boldsymbol{\theta}_i) \colon \Omega \to \mathbb{R}$ with parameters $\boldsymbol{\theta} = [\boldsymbol{\theta}_1, \dots, \boldsymbol{\theta}_N]$. Algorithm~\ref{alg:PI_PINN} trains these networks by extending Physics-Informed Policy Iteration~\cite{meng_physics-informed_2024} to $N$-player games while incorporating viscosity annealing~\cite{shilova_learning_2024}. Over $K_{max}$ iterations, Adam optimization minimizes the joint physics-informed objective~\eqref{eq:joint_pinn_loss}, with all spatial and temporal derivatives computed using automatic differentiation. Hereafter, we write $\hat{V}_i$ for brevity.

 \begin{remark}
     Since the exact viscosity solutions $V_i$ may exhibit gradient discontinuities, we augment \eqref{eq:HJB} with an artificial viscosity term $-\nu \Delta_{\mathbf{x}} V_i$ (see \eqref{eq:residual_1}), adapting the viscosity annealing method of \cite{shilova_learning_2024} from single-agent control to multiplayer games. We employ a diminishing viscosity schedule $\{\nu^{(k)}\}_{k=0}^K$ with $\nu^{(0)} = \nu_0 > 0$ and  $\nu^{(k)} \rightarrow 0$. For each $\nu > 0$, the viscous PDE admits classical $C^\infty(\Omega)$ solutions, ensuring that all required PINN derivatives are well-defined. By vanishing viscosity theory~\cite{bardi_optimal_1997}, the regularized solutions converge to the viscosity solution $V_i$ of \eqref{eq:HJB} as $\nu^{(k)} \to 0$.
\end{remark}

\begin{algorithm}
    \caption{$N$-Player PINN with Viscosity Annealing}\label{alg:PI_PINN}
    \begin{algorithmic}
        \State \textbf{Input}: Domain $\Omega$, dynamics $f$, $\{g_i\}$, and costs $\{r_i, \psi_i\}$
        \State \textbf{Hyperparameters}: Schedule $\{\nu^{(k)}\}_{k=0}^K$, $\lambda_T$,  $K_{max}$ 
        \State \textbf{Initialize}: Networks $\{\hat{V}_i(\cdot;\boldsymbol{\theta}_i^{(0)})\}$, $\nu \leftarrow \nu_0$, $k \leftarrow 0$ 
        \While{$k < K_{max}$} 
        \State Sample $\mathcal{D}_I \sim \mathrm{U}([t_0, T) \times \mathcal{X})$ and $\mathcal{D}_T  \sim \mathrm{U}(\mathcal{X})$
        \State \textbf{Policy Improvement}: 
        \State $\quad \hat{\mathbf{u}}_i^{(k+1)} \leftarrow \arg\min_{\hat{\mathbf{u}}_i} \Phi_i(\hat{\mathbf{u}}_i; \nabla_{\mathbf{x}}\hat{V}_i^{(k)})$ $\forall i \in \mathcal{N}$ \Comment{\eqref{eq:best_response}}
        \State \textbf{Policy Evaluation}: Minimize \eqref{eq:joint_pinn_loss} with frozen $\hat{\mathbf{u}}^{(k+1)}$
        \State Update $\nu \leftarrow \nu^{(k+1)}$ per schedule; $k \leftarrow k+1$ 
        \EndWhile 
        \State \textbf{Return:} $\{\hat{V}_i^{(K)}\}$ and induced strategy $\{\hat{\mathbf{u}}_i^{(K)}\}$    
    \end{algorithmic}
\end{algorithm}

In the policy improvement step, each player computes a best-response strategy via~\eqref{eq:best_response}. When available in closed form, this induced action is an explicit map of the value network derivatives. For example, control-affine dynamics and quadratic control costs yield, when $\mathcal{U}_i$ is inactive, $\hat{\mathbf{u}}_i^{(k)}(t, \mathbf{x}; \boldsymbol{\theta}_i^{(k)} ) = -\frac{1}{2} \mathbf{R}_i^{-1} g_i^\top \nabla_{\mathbf{x}} \hat{V}_i(t,\mathbf{x}; \boldsymbol{\theta}_i^{(k)})$.

During policy evaluation, the joint strategy profile $\hat{\pi}^{(k)}$ is frozen using stop-gradient operations, linearizing the policy evaluation subproblem for each player with respect to the unknown value function $\hat{V}_i$. For a frozen strategy profile $\hat{\pi}^{(k)}$ and viscosity level $\nu^{(k)} > 0$, player $i \in \mathcal{N}$ solves the resulting parabolic PDE by minimizing the residual 
\begin{multline}\label{eq:residual_1}
    L_{1, i}(t, \mathbf{x}; \boldsymbol{\theta}_i, \nu^{(k)}) = \partial_t  \hat{V}_i(t, \mathbf{x}; \boldsymbol{\theta}_i) \\
    + \Phi_i(t, \mathbf{x}, \hat{\mathbf{u}}_i
; \nabla_\mathbf{x} \hat{V}_i) - \nu^{(k)} \Delta_{\mathbf{x}} \hat{V}_i \,,
\end{multline}
where $\Delta_{\mathbf{x}}$ denotes the Laplacian, and the control $\hat{\mathbf{u}}_i^{(k)}$ is frozen with respect to $\boldsymbol{\theta}_i$. Setting $L_{1, i} = 0$ recovers the $\nu$-regularized policy evaluation PDE for player $i$, which admits smooth solutions for $\nu^{(k)} > 0$---crucial for the stability of Algorithm~\ref{alg:PI_PINN}. Further, the terminal residual
\begin{equation}\label{eq:residual_2} 
    L_{2, i}(\mathbf{x}; \boldsymbol{\theta}_i) = \hat{V}_i(T, \mathbf{x}; \boldsymbol{\theta}_i) - \psi_i(\mathbf{x}) \,, 
\end{equation} 
enforces the boundary condition $V_i(T, \mathbf{x}) = \psi_i(\mathbf{x})$ for the fixed-horizon game considered here.

The two residuals are evaluated at interior points $\mathcal{D}_I = \left\{(t^{(d)}, \mathbf{x}^{(d)})\right\}_{d=1}^{D_I} \subset [t_0, T) \times \mathcal{X} $ and terminal points $\mathcal{D}_T = \left\{\mathbf{x}^{(d)}\right\}_{d=1}^{D_T} \subset \mathcal{X}$, respectively. Both are uniformly resampled at each iteration to mitigate overfitting. While denser sampling may improve empirical training, the validity of the computable $\varepsilon$-Nash guarantees derived in Section~\ref{sec:results} is independent of the specific sampling scheme. 

Evaluated over these points, the residuals yield the per-player PINN objective under the frozen strategy profile $\hat{\pi}^{(k)}$
\begin{multline}\label{eq:pinn_loss}
    L_{PINN, i}(\boldsymbol{\theta}_i) = \frac{1}{D_I} \sum_{d  = 1}^{D_I} \left| L_{1, i}( t^{(d)}, \mathbf{x}^{(d)}; \boldsymbol{\theta}_i, \nu^{(k)}) \right|^2 \\
    + \frac{\lambda_T}{D_T} \sum_{d = 1}^{D_T}  \left| L_{2, i}(\mathbf{x}^{(d)}; \boldsymbol{\theta}_i) \right|^2 \,,
\end{multline}
where $\lambda_T > 0$ adaptively balances PDE and terminal residuals to stabilize training. Thus, the joint objective is 
\begin{equation}\label{eq:joint_pinn_loss}
    L_{PINNs}(\boldsymbol{\theta}) = \sum_{i \in \mathcal{N}} L_{PINN, i}(\boldsymbol{\theta}_i) \,.
\end{equation}

\section{Main Results}\label{sec:results}
We establish that our PINN-based PI framework yields a computable $\varepsilon$-Nash strategy profile for $N$-player differential games. We verify these guarantees a posteriori using $\partial$-CROWN certified residual bounds, which bound per-player regret under an approximate best-response condition. Therefore, the $\varepsilon$-Nash guarantee requires neither uniqueness nor explicit identification of viscosity solutions, which viscosity annealing approximates via $\nu$-regularization. We then prove a finite-time stopping criterion given small residuals and, finally, derive an explicit approximation error bound for two-player zero-sum games under Isaacs conditions.
 
\begin{lemma}[Representability]\label{lemma:representability}
    For any $\nu > 0$, let $V_i^\nu \in C^\infty(\Omega)$ solve the $\nu$-regularized HJB equation for player $i \in \mathcal{N}$. If the network class $\mathcal{V} = \{\hat{V}(\cdot; \boldsymbol{\theta})\}$ is dense in $C^{1}(\Omega)$, then for any $\delta > 0$, there exists $\boldsymbol{\theta}_{\delta}$ such that
    \begin{equation}
        \| \hat{V}_i(\boldsymbol{\theta}_\delta) - V_i^\nu \|_\infty \leq \delta \text{ and } \|\nabla\hat{V}_i(\boldsymbol{\theta}_{\delta}) - \nabla V_i^\nu \|_\infty \leq \delta \,. 
    \end{equation}
\end{lemma}

\begin{proof}
Since $V_i^\nu \in C^\infty(\Omega) \subset C^1(\Omega)$, the result follows immediately from the density of $\mathcal{V}$ in $C^1(\Omega)$ under $\|\cdot\|_{C^1(\Omega)}$. Hence, there exists $\boldsymbol{\theta}_{\delta}$ such that $\|\hat V_i(\boldsymbol{\theta}_{\delta})-V_i^\nu\|_{C^1(\Omega)} = \|\hat V_i(\boldsymbol{\theta}_{\delta}) - V_i^\nu\|_\infty + \|\nabla\hat V_i(\boldsymbol{\theta}_{\delta}) - \nabla V_i^\nu\|_\infty \leq \delta$.
\end{proof}

\begin{remark}
    The density assumption in Lemma~\ref{lemma:representability} is satisfied by standard neural network architectures. Deep networks with smooth, bounded nonpolynomial activations, such as tanh, are universal approximators of functions and their derivatives on compact sets~\cite{hornik_approximation_1991}. This is also true for deep ReLU and leaky-ReLU networks within Sobolev spaces~\cite{czarnecki_sobolev_2017}.
\end{remark}

\begin{lemma}[Residual Certification via $\partial$-CROWN]\label{lemma:worst-case_guarantee}
Let $\overline{\Omega} \subset [t_0, T] \times \mathbb{R}^n$ be a hyper-rectangle such that $\Omega \subseteq \overline{\Omega}$. Assume $\hat V_i(\cdot; \boldsymbol{\theta}_i^{(k)})$ belongs to a class supported by $\partial$-CROWN~\cite{eiras_efficient_2024}, and residuals \eqref{eq:residual_1}--\eqref{eq:residual_2} are explicit compositions of $\hat{V}_i$, $\partial_t \hat{V}_i$, $\nabla_{\mathbf{x}}\hat{V}_i$, and $\Delta_{\mathbf{x}}\hat{V}_i$ on $\overline\Omega$. Then, for every outer iterate $k$ and $i \in \mathcal{N}$, there exist computable $\delta_{1,i}^{(k)},\delta_{2,i}^{(k)}\geq 0$ such that
    \begin{equation*}
        \| L_{1, i}(t, \mathbf{x}; \boldsymbol{\theta}_i^{(k)}, \nu^{(k)}) \|_\infty \leq \delta_{1, i}^{(k)}\text{ and } \| L_{2, i}(\mathbf{x}; \boldsymbol{\theta}_i^{(k)}) \|_\infty \leq \delta_{2, i}^{(k)}.
    \end{equation*}
\end{lemma}

\begin{proof}
We directly apply the $\partial$-CROWN framework~\cite{eiras_efficient_2024}, which establishes guaranteed, deterministically computable worst-case upper and lower bounds on the residual errors (certification) of PINNs across a bounded continuous domain $\overline{\Omega}$. Define $\| \cdot \|_{\infty,\overline\Omega} = \sup_{\overline{\Omega}}|\cdot|$. As $\Omega\subseteq\overline\Omega$, it follows that
    \begin{equation}\label{eq:sup_restriction} 
    \| \cdot \|_\infty \leq \| \cdot \|_{\infty,\overline{\Omega}} \,. 
    \end{equation}
    
    Fix $i \in \mathcal{N}$ and a training iteration $k \in \{0,\dots,K\}$. Since the terminal residual \eqref{eq:residual_2} depends on zeroth-order network evaluations, we can bound $\hat{V}_i(\cdot; \boldsymbol{\theta}_i)$ using a subroutine of $\partial$-CROWN. Combined with known bounds of $\psi_i(\cdot)$, we obtain the upper and lower bounds $\alpha_{2, i}^{(k)} \leq L_{2, i}(\mathbf{x}; \boldsymbol{\theta}_i^{(k)}) \leq \beta_{2, i}^{(k)}$ for all $(t=T, \mathbf{x}) \in \overline{\Omega}$. Set $\delta_{2, i}^{(k)} = \max\{|\alpha_{2, i}^{(k)}|, |\beta_{2, i}^{(k)}|\}$ such that $ \| L_{2, i}(\mathbf{x}; \boldsymbol{\theta}_i^{(k)}) \|_{\infty, \overline{\Omega}} \leq \delta_{2, i}^{(k)}$. By \eqref{eq:sup_restriction}, we have that $ \| L_{2, i}(\mathbf{x}; \boldsymbol{\theta}_i^{(k)}) \|_\infty \leq \delta_{2, i}^{(k)}$ on the domain $\Omega$. 

    For the PDE residual, the assumed game dynamics make the induced actions $\hat{\mathbf{u}}_i$ explicit functions of the value network. For example, $\hat{\mathbf{u}}_i(t, \mathbf{x}) = -\frac{1}{2} \mathbf{R}_i^{-1} g_i^\top \nabla_\mathbf{x} \hat{V}_i(t, \mathbf{x}; \boldsymbol{\theta}_i)$. Hence, the PDE residual~\eqref{eq:residual_1} is an explicit composition of first- and second-order derivatives $\partial_t\hat V_i$, $\nabla_\mathbf{x}\hat V_i$ and $\Delta_{\mathbf{x}} \hat{V}_i$. Because these derivative operators and algebraic compositions are supported by the $\partial$-CROWN framework~\cite{eiras_efficient_2024}, the framework yields computable bounds $\alpha_{1, i}^{(k)} \leq L_{1, i}(t, \mathbf{x}; \boldsymbol{\theta}_i^{(k)}, \nu^{(k)}) \leq \beta_{1, i}^{(k)}$ for all $(t, \mathbf{x}) \in \overline{\Omega}$. Again, defining $\delta_{1, i}^{(k)} = \max\{|\alpha_{1, i}^{(k)}|, |\beta_{1, i}^{(k)}|\}$, we conclude that $ \| L_{1, i}(t, \mathbf{x}; \boldsymbol{\theta}_i^{(k)}, \nu^{(k)}) \|_\infty \leq \delta_{1, i}^{(k)}$ on the domain $\Omega$.
\end{proof}

\begin{remark}
    The $\partial$-CROWN~\cite{eiras_efficient_2024} framework supports architectures including, but not limited to, deep tanh, ReLU, and leaky-ReLU networks.
\end{remark}

To establish our computable equilibrium bound, we rely on the smoothness of the value networks. Since each $\hat{V}_i$ employs smooth activations, $\hat{V}_i \in C^\infty(\Omega) \subset C^1(\Omega)$. Therefore, the chain rule applies along any  absolutely continuous trajectory $\mathbf{x}(t)$ generated by measurable controls~\cite{fleming_deterministic_1975}, yielding
\begin{equation}\label{eq:total_derivative_of_value}
        \frac{d}{dt}\hat{V}(t, \mathbf{x}) = \partial_t \hat{V}(t, \mathbf{x}) + (\nabla_\mathbf{x} \hat{V}(t, \mathbf{x}))^\top \dot{\mathbf{x}} \,,
\end{equation}
for almost every $t \in [t_0, T]$ where $\dot{\mathbf{x}}(t)$ exists. 

\begin{proposition}[$\varepsilon$-Nash Bound for $N$-Player Differential Games]\label{proposition:equilibrium_guarantee}
    Consider an $N$-player game defined by \eqref{eq:generic_state_equation}--\eqref{eq:generic_cost_to_go}. Fix $\nu > 0$ and let the PINNs $\{\hat V_i\}_{i \in \mathcal{N}}$ be produced by Algorithm~\ref{alg:PI_PINN}. Assume there exist $\delta_{1, i}, \delta_{2, i}, \delta_{3, i} \geq 0$ such that for all $(t, \mathbf{x}) \in \Omega$ and $i \in \mathcal{N}$, each $\hat{V}_i$ satisfies the bounds $\|L_{1, i}(t, \mathbf{x}; \boldsymbol{\theta}_i, \nu)\|_\infty \leq \delta_{1, i}$ and $\|L_{2, i}(\mathbf{x}; \boldsymbol{\theta}_i)\|_\infty \leq \delta_{2, i} $, and the induced action profile $\hat{\mathbf{u}} = (\hat{\mathbf{u}}_1 \; \dots \; \hat{\mathbf{u}}_N)$ satisfies an approximate best-response condition 
\begin{equation}\label{assumption:best_response_gap}
    \Phi_i(t, \mathbf{x}, \hat{\mathbf{u}}_i;\hat{\mathbf{p}}_i) \leq \inf_{\mathbf{u}_i \in \mathcal{U}_i} \Phi_i(t, \mathbf{x}, \mathbf{u}_i;  \hat{\mathbf{p}}_i) + \delta_{3, i} \,,
\end{equation}
where $\hat{\mathbf{p}}_i = \nabla_{\mathbf{x}} \hat{V}_i(t, \mathbf{x})$. Then $\hat{\mathbf{u}}$ is an $\varepsilon$-Nash equilibrium with a per-player regret bound 
\begin{equation*} \varepsilon_i = (T - t_0)(2\delta_{1, i} + \delta_{3, i} + 2\nu M_{\Delta, i}) + 2\delta_{2, i} \,, 
\end{equation*} 
where $M_{\Delta, i} = \|\Delta_{\mathbf{x}} \hat{V}_i\|_\infty$ and $\varepsilon = \max_{i \in \mathcal{N}} \varepsilon_i$.
\end{proposition}

\begin{proof}
Recall that a profile $\pi^*$ is an $\varepsilon$-Nash equilibrium if there exists a scalar $\varepsilon \geq 0$ such that $J_i(t, \mathbf{x}, \pi_i^*, \pi_{-i}^*) \leq J_i(t, \mathbf{x}, \pi_i', \pi_{-i}^*) + \varepsilon$ for all  admissible $\pi_i' \in \Pi_i$. To show that under our residual and best-response assumptions no player can improve their cost by more than $\varepsilon_i$ through unilateral deviation, we fix $i \in \mathcal{N}$ and consider a trajectory  $\mathbf{x}^{\hat{\mathbf{u}}}$ generated by the induced profile $\hat{\mathbf{u}}$. By \eqref{eq:total_derivative_of_value} we have that for almost every $t \in [t_0, T]$
\begin{multline*}
        \frac{d}{dt}\hat{V}_i(t, \mathbf{x}^{\hat{\mathbf{u}}}) = \partial_t \hat{V}_i(t, \mathbf{x}^{\hat{\mathbf{u}}}) \\
        + \Phi_i(t, \mathbf{x}^{\hat{\mathbf{u}}}, \hat{\mathbf{u}}_i; \hat{\mathbf{p}}_i) - r_i(t, \mathbf{x}^{\hat{\mathbf{u}}},\hat{\mathbf{u}}_i, \hat{\mathbf{u}}_{-i})\,,
\end{multline*}
where $\hat{\mathbf{p}}_i = \nabla_\mathbf{x} \hat{V}_i$. Substituting the PDE residual \eqref{eq:residual_1} yields
\begin{equation*}
     \frac{d}{dt}\hat{V}_i(t, \mathbf{x}^{\hat{\mathbf{u}}}) = L_{1, i} - r_i(t, \mathbf{x}^{\hat{\mathbf{u}}},\hat{\mathbf{u}}_i, \hat{\mathbf{u}}_{-i}) +  \nu \Delta_{\mathbf{x}} \hat{V}_i   \,.
\end{equation*}
Integrating over the finite game horizon gives us
\begin{multline*}
    \int_{t_0}^T r_i(t, \mathbf{x}^{\hat{\mathbf{u}}},\hat{\mathbf{u}}_i, \hat{\mathbf{u}}_{-i}) \,dt = \hat{V}_i(t_0, \mathbf{x}_0) - \hat{V}_i(T, \mathbf{x}^{\hat{\mathbf{u}}}(T))\\
     + \int_{t_0}^T  \big(L_{1, i} + \nu \Delta_{\mathbf{x}} \hat{V}_i \big)dt \,.
\end{multline*}
Adding $\psi_i(\mathbf{x}^{\hat{\mathbf{u}}}(T))$ to both sides gives us an expression in terms of cost-to-go~\eqref{eq:generic_cost_to_go}
\begin{multline*}
     J_i(t_0, \mathbf{x}_0,\hat{\mathbf{u}}_i, \hat{\mathbf{u}}_{-i}) = \hat{V}_i(t_0, \mathbf{x}_0) - \hat{V}_i(T, \mathbf{x}^{\hat{\mathbf{u}}}(T))\\
      + \psi_i(\mathbf{x}^{\hat{\mathbf{u}}}(T)) + \int_{t_0}^T  \big(L_{1, i} + \nu \Delta_{\mathbf{x}} \hat{V}_i \big)dt \,.
\end{multline*}
Applying the uniform bounds $\|L_{1,i}\|_\infty \leq \delta_{1,i}$, $\|L_{2,i}\|_\infty \leq \delta_{2,i}$, and $\|\Delta_{\mathbf{x}} \hat{V}_i\|_\infty \leq M_{\Delta,i}$ gives us 
\begin{multline}\label{eq:proposition_1_upper_bound}
     J_i(t_0, \mathbf{x}_0,\hat{\mathbf{u}}_i, \hat{\mathbf{u}}_{-i}) \leq \hat{V}_i(t_0, \mathbf{x}_0) + \delta_{2, i} \\
     + (T - t_0)(\delta_{1, i} + \nu M_{\Delta, i}) \,.
\end{multline}

Consider a trajectory $\mathbf{x}^{\mathbf{u}'}$ generated from an arbitrary deviation $\mathbf{u}_i' \in \mathcal{U}_i$ by player $i$ with all other players playing $\hat{\mathbf{u}}_{-i}$ as before. By the approximate best-response condition \eqref{assumption:best_response_gap}, we have that $\Phi_i(t, \mathbf{x}^{\mathbf{u}'}, \mathbf{u}_i'; \hat{\mathbf{p}}_i) \geq \inf_{\mathbf{u}_i \in \mathcal{U}_i} \Phi_i(t, \mathbf{x}^{\mathbf{u}'}, \mathbf{u}_i; \hat{\mathbf{p}}_i) \geq \Phi_i(t, \mathbf{x}^{\mathbf{u}'}, \hat{\mathbf{u}}_i; \hat{\mathbf{p}}_i) - \delta_{3, i}$. Under this deviation, we have that
\begin{multline*} 
    \frac{d}{dt}\hat{V}_i(t, \mathbf{x}^{\mathbf{u}'}) = \partial_t \hat{V}_i(t, \mathbf{x}^{\mathbf{u}'}) \\+ \Phi_i(t, \mathbf{x}^{\mathbf{u}'}, \mathbf{u}_i'; \hat{\mathbf{p}}_i) - r_i(t, \mathbf{x}^{\mathbf{u}'}, \mathbf{u}_i', \hat{\mathbf{u}}_{-i}) \,.
    \end{multline*}
Substituting the residual evaluated at $(t, \mathbf{x}^{\mathbf{u}'}(t))$ then gives
\begin{multline*} 
\frac{d}{dt}\hat{V}_i(t, \mathbf{x}^{\mathbf{u}'}) = L_{1,i} - \Phi_i(t, \mathbf{x}^{\mathbf{u}'}, \hat{\mathbf{u}}_i; \hat{\mathbf{p}}_i) + \nu \Delta_{\mathbf{x}} \hat{V}_i \\ + \Phi_i(t, \mathbf{x}^{\mathbf{u}'}, \mathbf{u}_i'; \hat{\mathbf{p}}_i) - r_i(t, \mathbf{x}^{\mathbf{u}'}, \mathbf{u}_i', \hat{\mathbf{u}}_{-i}) \,. 
\end{multline*}
Following the same steps as before, we can apply the approximate best-response condition before integrating over the finite horizon and using uniform bounds to yield 
\begin{multline}\label{eq:proposition_1_lower_bound}
      J_i(t_0, \mathbf{x}_0, \mathbf{u}_i', \hat{\mathbf{u}}_{-i})  \geq \hat{V}_i(t_0, \mathbf{x}_0) - \delta_{2, i} \\
      - (T - t_0)(\delta_{1, i} + \delta_{3, i} + \nu M_{\Delta, i}) \,.
\end{multline}

    Subtracting \eqref{eq:proposition_1_lower_bound} from \eqref{eq:proposition_1_upper_bound} verifies
\begin{equation*}
   J_i(t_0, \mathbf{x}_0, \hat{\mathbf{u}}_i, \hat{\mathbf{u}}_{-i}) - J_i(t_0, \mathbf{x}_0, \mathbf{u}_i', \hat{\mathbf{u}}_{-i}) \leq  \varepsilon_i \,,
\end{equation*}
for $\varepsilon_i = (T - t_0)(2\delta_{1, i} + \delta_{3, i} + 2\nu M_{\Delta, i}) + 2\delta_{2, i}$. Define $\varepsilon = \max_{i \in \mathcal{N}}\varepsilon_i$. Thus, $\hat{\mathbf{u}}$ is an $\varepsilon$-Nash equilibrium, with $\varepsilon \to 0$ as $\delta_{1,i}, \delta_{2,i}, \delta_{3,i}, \nu M_{\Delta,i} \to 0$.
\end{proof}

This result demonstrates the linear scaling of $\varepsilon_i$ with the horizon length $T - t_0$, with the $\nu$-regularization term $\nu M_{\Delta,i}$ vanishing as $\nu^{(k)} \rightarrow 0$. Since residuals inherently increase with domain volume ($\overline{\Omega}$), achieving a fixed tolerance $\varepsilon$ requires certification such that $\delta_{1,i} = \mathcal{O}((T - t_0)^{-1})$.

\begin{corollary}\label{corollary:equilibrium_guarantee_PINN}
    In Algorithm~\ref{alg:PI_PINN}, the profile is the exact pointwise minimizer $\hat{\mathbf{u}}_i(t, \mathbf{x}) = \mathbf{u}_i^{br}(t, \mathbf{x}; \hat{\mathbf{p}}_i)$. Thus, Proposition~\ref{proposition:equilibrium_guarantee} holds with $\delta_{3, i} = 0$. Otherwise, $\delta_{3, i}$ captures the optimization and approximation error of player $i$'s actions.
\end{corollary}

\begin{proposition}[Finite Time Stopping of Algorithm~\ref{alg:PI_PINN}]\label{proposition:finite_time_convergence_N_player}
    Consider an $N$-player game \eqref{eq:generic_state_equation}--\eqref{eq:generic_cost_to_go} on a compact domain $\Omega$ and a hyper-rectangle $\overline{\Omega} \subset [t_0, T] \times \mathbb{R}^n$ such that $\Omega \subseteq \overline{\Omega}$. Let $\{\hat V_i(\cdot; \boldsymbol{\theta}_i^{(k)})\}_{i\in\mathcal N}$ be a set of PINNs produced by Algorithm~\ref{alg:PI_PINN}. Assume that Lemma~\ref{lemma:worst-case_guarantee} holds for each outer iterate $k\le K$, yielding uniform bounds $\delta_{1,i}^{(k)},\delta_{2,i}^{(k)} \geq 0$. For a target tolerance $\varepsilon > 0$, define the stopping iteration
    \begin{multline}\label{eq:stopping_time}
        k_\varepsilon = \inf\Big\{k \leq K:\; \max_{i \in \mathcal{N}} \Big[ 
    (T-t_0)\big(2\delta_{1,i}^{(k)} \\+ 2\nu^{(k)} M_{\Delta, i}^{(k)}\big)+2\delta_{2,i}^{(k)} \Big] \le \varepsilon\Big\} \,.
    \end{multline}
If this set is nonempty, then $k_\varepsilon < \infty$ and the strategy profile induced by $\{ \hat V_i(\cdot; \boldsymbol{\theta}_i^{(k_\varepsilon)}) \}_{i\in\mathcal N}$ is an $\varepsilon$-Nash equilibrium. 
\end{proposition}

\begin{proof}
    To prove that Algorithm~\ref{alg:PI_PINN} reaches an $\varepsilon$-Nash equilibrium in finite time, consider that each iteration $k$ involves computing the exact best response $\mathbf{u}_i^{(k+1)}(t,\mathbf{x}) = \mathbf{u}_i^{br}(t, \mathbf{x}; \nabla_{\mathbf{x}} \hat V_i^{(k)}(t, \mathbf{x}))$ on $\overline\Omega$. By Corollary~\ref{corollary:equilibrium_guarantee_PINN} we have that $\delta_{3, i}^{(k)} = 0$ for all $i \in \mathcal{N}$ and all $k$. 

Then by Lemma~\ref{lemma:worst-case_guarantee} and exact best responses ($\delta_{3,i}^{(k)}=0$), Proposition~\ref{proposition:equilibrium_guarantee} gives us 
\begin{equation*}
    \varepsilon_i^{(k)} \le (T-t_0)(2\delta_{1,i}^{(k)}+2\nu^{(k)}M_{\Delta,i}^{(k)})+2\delta_{2,i}^{(k)} \,,
\end{equation*}
for all $i \in \mathcal{N}$. Thus, the stopping time $k_\varepsilon$ is well defined, guaranteeing an $\varepsilon$-Nash equilibrium.

    Choose $\eta = \frac{\varepsilon}{4(T - t_0) + 2}$. As is standard in the PINN literature~\cite{raissi_physics_2019, eiras_efficient_2024}, assume there exists a finite \(K_\eta < \infty\) such that for all \(k \geq K_\eta\), $\max_{i\in\mathcal{N}} \{\delta_{1,i}^{(k)}, \delta_{2,i}^{(k)}, \nu^{(k)} M_{\Delta, i}^{(k)}\} \leq \eta$. Therefore, for all $k \geq K_\eta$, the left side of \eqref{eq:stopping_time} satisfies
\begin{multline*} 
    \max_{i \in \mathcal{N}} \big[(T-t_0)(2\delta_{1,i}^{(k)} + 2\nu^{(k)} M_{\Delta, i}^{(k)}) + 2\delta_{2,i}^{(k)}\big] \\ \leq (T-t_0)(2\eta + 2\eta) + 2\eta = 4(T-t_0)\eta + 2\eta = \varepsilon \,. 
\end{multline*}
Thus, the stopping set is nonempty and $k_\varepsilon \leq K_\eta < \infty$.       
\end{proof}    
Propositions~\ref{proposition:equilibrium_guarantee}--\ref{proposition:finite_time_convergence_N_player} provide sufficient conditions for computing $\varepsilon$-Nash bounds for PINN-derived solutions to $N$-player differential games. These guarantees are conditional on sufficiently small residuals, a standard assumption in PINN analysis~\cite{meng_physics-informed_2024}, and the availability of domain-wide certified residual bounds (Lemma~\ref{lemma:worst-case_guarantee}) over $\Omega$. Lemma~\ref{lemma:representability} supports this framework by showing that the $\nu$-regularized solution can be approximated arbitrarily well within the chosen network class.

In practice, tight certification over the full domain can be difficult to achieve, particularly in adversarial nonzero-sum games, where exploitation of local approximation errors can loosen the equilibrium bound. This motivates the next analysis for zero-sum games, where the comparison principle enables a stronger pointwise value-function guarantee, explicitly quantifying how PDE residual errors propagate to solutions.

\begin{proposition}[Approximation Error in Two-Player Zero-Sum Games]\label{proposition:2PZS_approximation}
    Consider a two-player zero-sum differential game on $\Omega$ satisfying the Isaacs condition~\cite{basar_dynamic_1999} and a viscosity comparison principle~\cite{bardi_optimal_1997} such that the value $V^*$ is the unique viscosity solution of 
\begin{equation}
    \partial_t V^*(t, \mathbf{x})+\mathcal{H}(t, \mathbf{x},\nabla_{\mathbf{x}} V^*(t, \mathbf{x}))=0 \,,
\end{equation}
    subject to $V^*(T, \mathbf{x})=\psi(\mathbf{x})$ where $\mathcal{H}(t,\mathbf{x},\mathbf{p})=\min_{\mathbf{u}_1}\max_{\mathbf{u}_2}H(t,\mathbf{x},\mathbf{p},\mathbf{u}_1,\mathbf{u}_2)$. Let $\hat V$ be a PINN produced by Algorithm~\ref{alg:PI_PINN} with a final $\nu > 0$. Assume there exist $\delta_{1}, \delta_{2} \geq 0$ such that for all $(t, \mathbf{x}) \in \Omega$, $\hat{V}$ satisfies 
\begin{equation}\label{assumption:residuals}
    \|L_{1}(t, \mathbf{x}; \boldsymbol{\theta}, \nu)\|_\infty \leq \delta_{1} \text{ and } \|L_{2}(\mathbf{x}; \boldsymbol{\theta})\|_\infty \leq \delta_{2} \,.
\end{equation}

\emph{A posteriori} approximation error is bounded in terms of $M_\Delta = \| \Delta_{\mathbf{x}} \hat{V} \|_\infty$ (as calculated via neural network verification) by
    \begin{equation}
            \| \hat{V} - V^* \|_\infty \leq (T - t_0)(\delta_1 + \nu M_{\Delta}) + \delta_2 \,.
    \end{equation}
\end{proposition}

\begin{proof}
     We decompose the approximation error via the triangle inequality as $\|\hat{V} - V^*\|_\infty \leq \|\hat{V} - V^\nu\|_\infty + \|V^\nu - V^*\|_\infty$, where $V^\nu$ denotes the unique $C^\infty(\Omega)$ solution to the $\nu$-regularized HJB equation
\begin{equation}\label{eq:viscous_HJB_2PZS}
    \partial_t V^\nu + \mathcal{H}(t, \mathbf{x}, \nabla_{\mathbf{x}} V^\nu) - \nu \Delta_{\mathbf{x}} V^\nu = 0 \,,
\end{equation}
subject to $V^\nu(T, \mathbf{x}) = \psi(\mathbf{x})$.

    Define envelope functions $v^\pm(t, \mathbf{x}) = \hat{V}(t, \mathbf{x}) \pm [\delta_1 (T - t) + \delta_2]$. We verify that $v^+$ is a viscosity supersolution at any $(t, \mathbf{x}) \in \Omega$. Since $\partial_t v^+ = \partial_t \hat{V} - \delta_1$, $\nabla_{\mathbf{x}} v^+ = \nabla_{\mathbf{x}} \hat{V}$, and $\Delta_{\mathbf{x}}v^+ = \Delta_{\mathbf{x}} \hat{V}$ we compute
\begin{align*}
    \partial_t v^+ &+ \mathcal{H} (t, \mathbf{x}, \nabla_{\mathbf{x}} v^+) -  \nu \Delta_{\mathbf{x}}v^+\\
    &= \partial_t \hat{V} - \delta_1 + \mathcal{H}(t, \mathbf{x}, \nabla_{\mathbf{x}} \hat{V}) -  \nu \Delta_{\mathbf{x}} \hat{V}\\
    &\leq \delta_1 - \delta_1 = 0 \,.
\end{align*}
Similarly, $v^-$ is a viscosity subsolution satisfying
 \begin{align*}
    \partial_t v^- &+ \mathcal{H} (t, \mathbf{x}, \nabla_{\mathbf{x}} v^-)  -  \nu \Delta_{\mathbf{x}}v^-\\
    &= \partial_t \hat{V} + \delta_1 + \mathcal{H}(t, \mathbf{x}, \nabla_{\mathbf{x}} \hat{V})  -  \nu \Delta_{\mathbf{x}}\hat{V}\\
    &\geq \delta_1 - \delta_1 = 0 \,.
\end{align*}
   
   At $t=T$ the residual assumption implies that $\hat{V}(T, \mathbf{x}) - \delta_2 \leq \psi(\mathbf{x}) \leq \hat{V}(T, \mathbf{x}) + \delta_2$ for all $\mathbf{x} \in \mathcal{X}$. By definition of $v^\pm$, this establishes the boundary ordering $v^-(T, \mathbf{x}) \leq V^\nu(T, \mathbf{x}) \leq v^+(T, \mathbf{x})$. Since the Isaacs condition holds, the comparison principle guarantees that this ordering is preserved over the entire domain $\Omega$ such that $v^-(t, \mathbf{x}) \leq V^\nu(t, \mathbf{x}) \leq v^+(t, \mathbf{x})$~\cite{bardi_optimal_1997}. Substituting $v^\pm$ then yields
\begin{equation*} 
   \| \hat{V}(t, \mathbf{x}) - V^\nu(t, \mathbf{x}) \|_\infty \leq \delta_1(T - t) + \delta_2 \leq \delta_1(T - t_0) + \delta_2 \,. 
\end{equation*}

    Letting  $w = V^\nu - V^*$, the error between the $\nu$-regularized solution and $V^*$ can be expressed as 
    \begin{equation*}
        \partial_t w + [ \mathcal{H}(t, \mathbf{x}, \nabla_{\mathbf{x}} V^\nu) -  \mathcal{H}(t, \mathbf{x}, \nabla_{\mathbf{x}} V^*)] - \nu\Delta_{\mathbf{x}} V^\nu = 0\,,
    \end{equation*}
subject to $w(T, \mathbf{x}) = 0$. Construct envelope functions $w^\pm = \pm \nu M_{\Delta}(T - t)$ where $M_\Delta = \| \Delta_\mathbf{x} V^\nu \|_\infty$ or  $M_\Delta \approx \|\Delta_{\mathbf{x}} \hat{V}\|_\infty$ as a computable proxy. Therefore, $\partial_t w^\pm = \mp \nu M_\Delta$ and $\nabla_{\mathbf{x}} w^\pm = 0$. We verify that $w^+$ is a viscosity supersolution. Spatial constancy implies $\nabla_{\mathbf{x}}(V^* + w^+) = \nabla_{\mathbf{x}} V^*$. Thus,
\begin{align*} 
\partial_t w^+ &+ [\mathcal{H}(\cdots, \nabla_{\mathbf{x}} V^\nu) - \mathcal{H}(\cdots, \nabla_{\mathbf{x}} V^*)] - \nu \Delta_{\mathbf{x}} V^\nu \\
&= -\nu M_\Delta + 0 - \nu \Delta_{\mathbf{x}} V^\nu \\ 
&\leq -\nu M_\Delta + \nu M_\Delta = 0 \,, 
\end{align*} 
where $-\nu\Delta_{\mathbf{x}} V^\nu \leq \nu M_\Delta$ by definition. Similarly, it follows that $w^-$ is a subsolution. At $t=T$, we have $w^\pm(T, \mathbf{x}) = 0 = w(T, \mathbf{x})$. By the comparison principle we have that $w^- \leq w \leq w^+$ on $\Omega$. Therefore,
\begin{equation*} 
   \| V^\nu(t, \mathbf{x}) - V^*(t, \mathbf{x}) \|_\infty \leq \nu M_{\Delta}(T - t) \leq \nu M_{\Delta}(T - t_0) \,. 
\end{equation*}
    Applying the triangle inequality completes the proof.
\end{proof}

\section{Numerical Example}\label{sec:experiments}
Consider an $N$-player pursuit-evasion game comprising $N_p$ pursuers and $N_e$ evaders, indexed by $\mathcal{N}_p$ and $\mathcal{N}_e$, respectively. We consider two environments, both with linear dynamics of the form
\begin{equation}\label{eq:MPPE_state_equation} 
    \dot{\mathbf{x}} = \mathbf{A}\mathbf{x} + \sum_{i \in \mathcal{N}} \mathbf{B}_i \mathbf{u}_i \,.
\end{equation} 
We demonstrate analytical validation (\mbox{$N=2$}) and multiplayer scalability (\mbox{$N=4$}) over the horizon $[0, 10.0]$. PINNs are trained via Algorithm~\ref{alg:PI_PINN} using uniformly sampled interior ($D_I$) and terminal ($D_T$) points in $8$ minibatches. For \mbox{$N=2$}, the game is zero-sum, reducing certification to one shared value network ($2 \times 64$ neurons) trained with \mbox{$K_{max}=1{,}500$}, \mbox{$D_I=1{,}792$}, and \mbox{$D_T=256$}. For \mbox{$N=4$}, each per-player network ($3 \times 128$ neurons) is trained with \mbox{$K_{max}=1{,}800$}, \mbox{$D_I=31{,}130$}, and \mbox{$D_T=1{,}638$}.

\textbf{Computational scaling}: $20{,}000$-branch greedy input splitting is used for certification bounds ($\delta_{1, i}$, $\delta_{2, i}$, and $M_{\Delta,i}$). Offline $\partial$-CROWN scales linearly with player count $N$ and state dimension $n$, and polynomially with network size~\cite{eiras_efficient_2024}.

\textbf{Analytical validation ($\Omega \subset \mathbb{R}^{4}$)}: One pursuer chases one evader, both moving according to first-order velocity-limited kinematics in 3D space. With $\mathbf{A} = \mathbf{0}$, $\mathbf{B}_p = -\mathbf{B}_e = \mathbf{I}$, velocity limits $\|\mathbf{u}_i\| \leq u_{max,i}$, and terminal cost $\psi = \|\mathbf{x}\|^2$ in relative coordinates $\mathcal{X} \subset \mathbb{R}^3$, the analytical solution is $V^* = \max\{0, \|\mathbf{x}\| - (u_{max,p} - u_{max,e})(T-t)\}^2$ with bang-bang controls $\mathbf{u}_i^* = -u_{max,i} \cdot (\nabla V/\|\nabla V\|)$. 

\begin{figure}
    \centering
    \vspace{2.5mm}
    \includegraphics[width=0.97\linewidth]{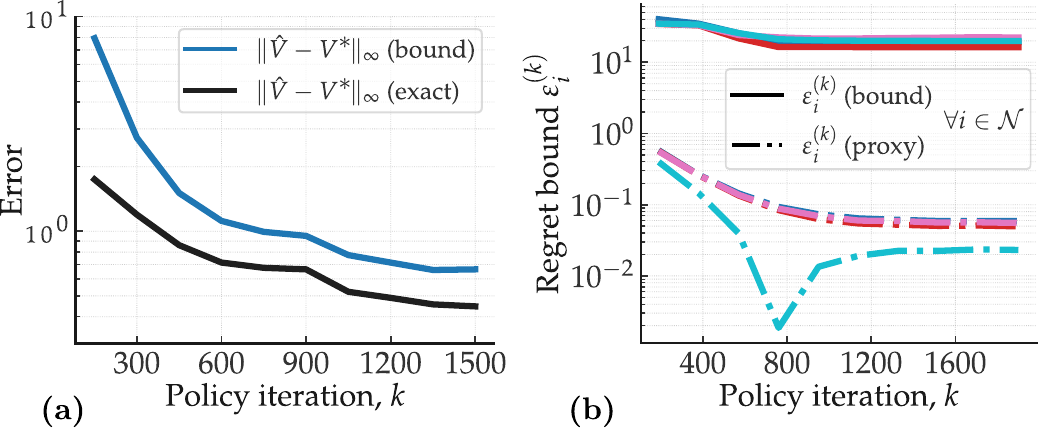}
    \caption{Convergence: \textbf{(a)} Pointwise error bound  (Proposition~\ref{proposition:2PZS_approximation}) vs. true analytical error; \textbf{(b)} Computable worst-case per-player regret $\varepsilon_i$ (Proposition~\ref{proposition:equilibrium_guarantee}) vs. $L^2$ training proxy.}
    \label{fig:convergence}
\end{figure}

\begin{figure}
     \centering
     \includegraphics[width=0.97\linewidth]{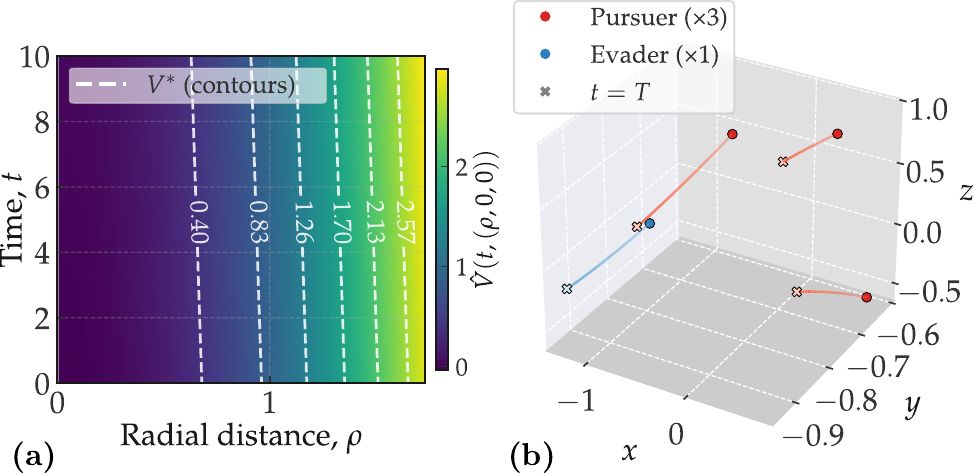}
     \caption{Value approximation: \textbf{(a)} $\hat{V}$ surface (color) vs. $V^*$ contours (white dashed) at states $\mathbf{x} = (\rho, 0, 0)$, consistent with the bound $\|\hat{V} - V^*\|_\infty \lesssim 0.66$; \textbf{(b)} Per-player trajectories for $N=4$.}
     \label{fig:behaviour}
\end{figure}

Algorithm~\ref{alg:PI_PINN} achieves $\delta_1^{(1500)} \approx 2.3\times10^{-2}$ and $\delta_2^{(1500)} \approx 0.43$, guaranteeing $\|\hat{V} - V^*\|_\infty \lesssim 0.66$ and $\varepsilon \lesssim 1.3$ ($\varepsilon/T \lesssim 0.13$) over $T=10.0$ (Fig.~\ref{fig:convergence}(a)). The monotonic convergence of the bound (Fig.~\ref{fig:convergence}(a)) and tight $\hat{V}$ surface alignment with $V^*$ contours (Fig.~\ref{fig:behaviour}(a)) validate the approximation bound.

\textbf{Multiplayer scalability ($\Omega \subset \mathbb{R}^{25}$)}: For $N_p = 3$ pursuers and $N_e = 1$ evader with second-order acceleration-limited kinematics and quadratic costs, the finite-horizon Nash equilibrium---when it exists---admits quadratic value functions~\cite{basar_dynamic_1999}, but grid-based computation becomes intractable. For $\mathcal{X} \subset \mathbb{R}^{24}$, we select a drag coefficient $\gamma \in [0, 1]$ and appropriately stack
\begin{equation}
    \mathbf{A}_i = 
\begin{bmatrix}
    0 & \mathbf{I} \\
    0 & -\gamma\mathbf{I}
\end{bmatrix} \text{ and } \mathbf{B}_i = \begin{bmatrix}
    0\\ \mathbf{I}
\end{bmatrix} \,.
\end{equation}
Each player $i$ has quadratic cost $J_i = \int_0^T (\mathbf{u}_i^\top \mathbf{R}_i \mathbf{u}_i + \mathbf{x}^\top \mathbf{Q}_i \mathbf{x} ) dt + \mathbf{x}^\top \mathbf{F}_i \mathbf{x}|_{t=T}$, where $\mathbf{Q}_i$ and $\mathbf{F}_i$ encode team objectives.

Algorithm~\ref{alg:PI_PINN} achieves a conservative \emph{a posteriori} guarantee $\varepsilon \approx 2.2\times 10^1$ ($\varepsilon/T \approx 2.2$) over the full kinematic domain and $T = 10.0$ in $1000$ iterations. The $L^2$ training proxy converges to  $\hat{\varepsilon} \approx 5.9\times 10^{-2}$, confirming approximation quality (Fig.~\ref{fig:convergence}(b)). Induced strategies exhibit coordinated pursuit-evasion behaviors illustrated in Fig.~\ref{fig:behaviour}(b).

\section{Conclusions \& Future Work} 
We presented a PINN-based PI framework for $N$-player nonzero-sum differential games, combining the structural stability of PI with viscosity annealing to derive computable $\varepsilon$-Nash and finite-time stopping guarantees. This demonstrates that physics-informed learning effectively enforces game-theoretic consistency in high-dimensional differential games. Future work will extend our framework to handle hard state constraints via boundary-residual formulations and explore using these PDE residuals as structural priors to regularize multi-agent RL for games with unknown dynamics.


\bibliographystyle{ieeetr}
\bibliography{arxiv}
\end{document}